\documentclass{aa}
\usepackage{graphicx,times}

\newcommand{\xmm}{{\it XMM-Newton}}
\newcommand{\chandra}{{\it Chandra}}
\newcommand{\rgs}{{RGS}}
\newcommand{\epic}{EPIC}

\newcommand{\pn}{EPIC-pn}

\newcommand{\dem}{DEM L\,71}

\begin{document}

\title{High resolution spectroscopy and emission line imaging of \dem\ with 
\xmm}

\author{K.J. van der Heyden\inst{1},
 J.A.M. Bleeker\inst{1},
 J.S. Kaastra\inst{1},
 J. Vink\inst{2,3}}
 
\offprints{K.J. van der Heyden}

\institute{SRON National Institute for Space Research, 
Sorbonnelaan 2, 
3584 CA Utrecht, The Netherlands \and 
Columbia Astrophysics Laboratory, Columbia University, 550 West 1
20th Street, New York, NY 10027, USA 
\and 
\chandra\ fellow 
\\email: K.J.van.der.Heyden@sron.nl \\} 
 
\titlerunning{\xmm\ observations of \dem} 
\authorrunning{K.J. van der Heyden et al.} 
 
\date{Received 5 February 2003 / Accepted 15 April 2003} 
 
\abstract{The X-ray emission from the supernova remnant \dem\ is measured
and analysed using the high-resolution cameras and spectrometers on board \xmm.
The spectrum from the outer shell is reproduced very well by two plasma
components of $kT_{\rm e}$ = 0.3 and 0.8~keV. The abundance value from this
shell is consistent with the average LMC values. More extreme temperature
variations are possibly indicated by spatial variations in the \ion{O}{vii} 
forbidden/resonance line ratio, which could imply that in some regions the 
plasma is cooling dramatically and recombining. However, an alternative and 
equally interesting possibility is that the variation in forbidden/resonance 
ratios is due to resonant scattering, which would reduce resonance line 
emission along lines of sight with a high \ion{O}{vii} column density. The 
inner region is hotter ($kT_{\rm e}$ = 1.1~keV) and shows enhanced Fe and Si 
abundances. We estimate the Fe and Si ejecta mass to be 0.7--1.1 M$_{\sun}$ and 
0.1--0.15 M$_{\sun}$, respectively. The morphology, mass estimates and 
abundances strongly suggest that \dem\ is the result of a type Ia explosion, as 
indicated by previous measurements.

}

\maketitle

\section{Introduction} 

\dem\ (SNR 0505-67.9) belongs to a class of Supernova Remnants (SNR) that
are dominated by hydrogen emission in their optical spectra with virtually no
emission from collisionally excited forbidden lines. Analysis of the optical
spectrum of \dem\ indicates a shock velocity of 300--800\ km\,s$^{-1}$ which
corresponds to a shock temperature of $T_{\rm s}=0.11-0.75$\ keV (Smith et al.
\cite{smith}). This information, combined with a radial size of $\sim$ 10 pc,
gives an estimated age of $\sim$ 10$^{4}$\ years. \textit{ASCA} spectroscopy
revealed that, despite its age, \dem\ still showed emission from ejecta
material in the form of enhanced Fe abundance (Hughes et al. \cite{hughes2}).
The enhanced Fe abundance indicates that \dem\ is the result of a type Ia SNe.
The recent \chandra\ Fe-L images (Hughes et al. \cite{hughes03}; see 
also Fig.~\ref{fig:pn_im}) also reveal 
that the remnant exhibits both shell-like X-ray emission and a centrally filled
morphology in this energy range. 

In this work, we present the X-ray spectra of \dem\ measured by the
\xmm\ suite of scientific instruments, i.e. the Reflection Grating
Spectrometers (\rgs) (den Herder et al. \cite{denherder}) and the European 
Photon Imaging Cameras (\epic) (Turner et al. \cite{turner}; Str{\"u}der et 
al. \cite{struder}).
In particular, we exploit the high dispersion of the \rgs\ and its unique
capability to resolve line emission from extended sources to investigate
the conditions of the X-ray emitting plasma in \dem\ in unprecedented
detail.

\section{Observation and Reduction}

Our analysis is based on \xmm\ observations of \dem\ obtained on 2
April 2001. The observation was performed with the telescope rolled
such that the \rgs\ dispersion axis was aligned at 72\degr, clockwise on
the sky from celestial North. A large fraction of the observation is
affected by high particle background caused by solar activity. For the
\epic\ instruments periods of high particle background were rejected
based upon the 10--12~keV count rate for the entire field of view.
After filtering out these high background periods we are left with a net
exposure time of 10 ks for the \epic\ instruments. The \epic\ instruments were
all operated in full window mode with the medium filter in place. Good
time intervals (GTI) for the \rgs\ were created based on the count rate
in CCD number 9 (it is the one closest to the optical axis of the
telescope, therefore the most affected by background flares). After
applying the GTI filter we are left with an effective exposure time of
40 ks for the \rgs. The spectra were extracted by applying spatial
filters to the CCD image while a CCD pulse-height filter was applied to
select the $m=-1$ spectral order. All the \xmm\ data were processed
using the Science Analysis System (SAS version 5.3).

\section{Analysis and Results}

\begin{figure}
\resizebox{\hsize}{!}{\includegraphics{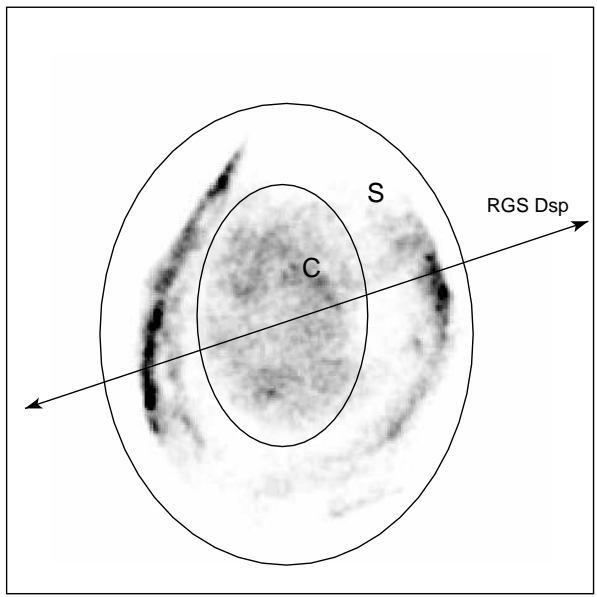}}
\caption[]{\chandra\ image for the approximate energy range 0.7--1.0 keV. The 
regions used for the extraction of the outer shell (S) and central region (C) 
are indicated. The alignment of the \rgs\ dispersion axis is also indicated. } 
\label{fig:pn_im} 
\end{figure}

\subsection{Spectra}

The spectral analysis was performed using the SRON SPEX package (Kaastra
et al.~\cite{kaastra}), which contains the MEKAL atomic database (Mewe
et al.~\cite{mewe}) for thermal emission. Our model for the X-ray
emission of \dem\ consists of several plasma components. For each
component, we fit for the volume emission measure ($n_{\rm e}n_{\rm
H}V$), the electron temperature ($T_{\rm e}$), the ionisation age
($n_{\rm e}t$) (Kaastra \& Jansen~\cite{net}), the elemental abundances,
the redshift of the source, and the column density $N_{\rm H}$ of
absorbing gas along the line of sight. Here, $n_{\rm e}$ is the
electron density, $n_{\rm H}$ is the hydrogen density, $V$ is the volume
of emitting gas and $t$ is the time since the material has been shocked
and heated to its current temperature. A distance of 52 kpc to the LMC is 
assumed (Feast et al.~\cite{feast}).

\begin{figure*}
\resizebox{\hsize}{!}{\includegraphics[angle=-90]{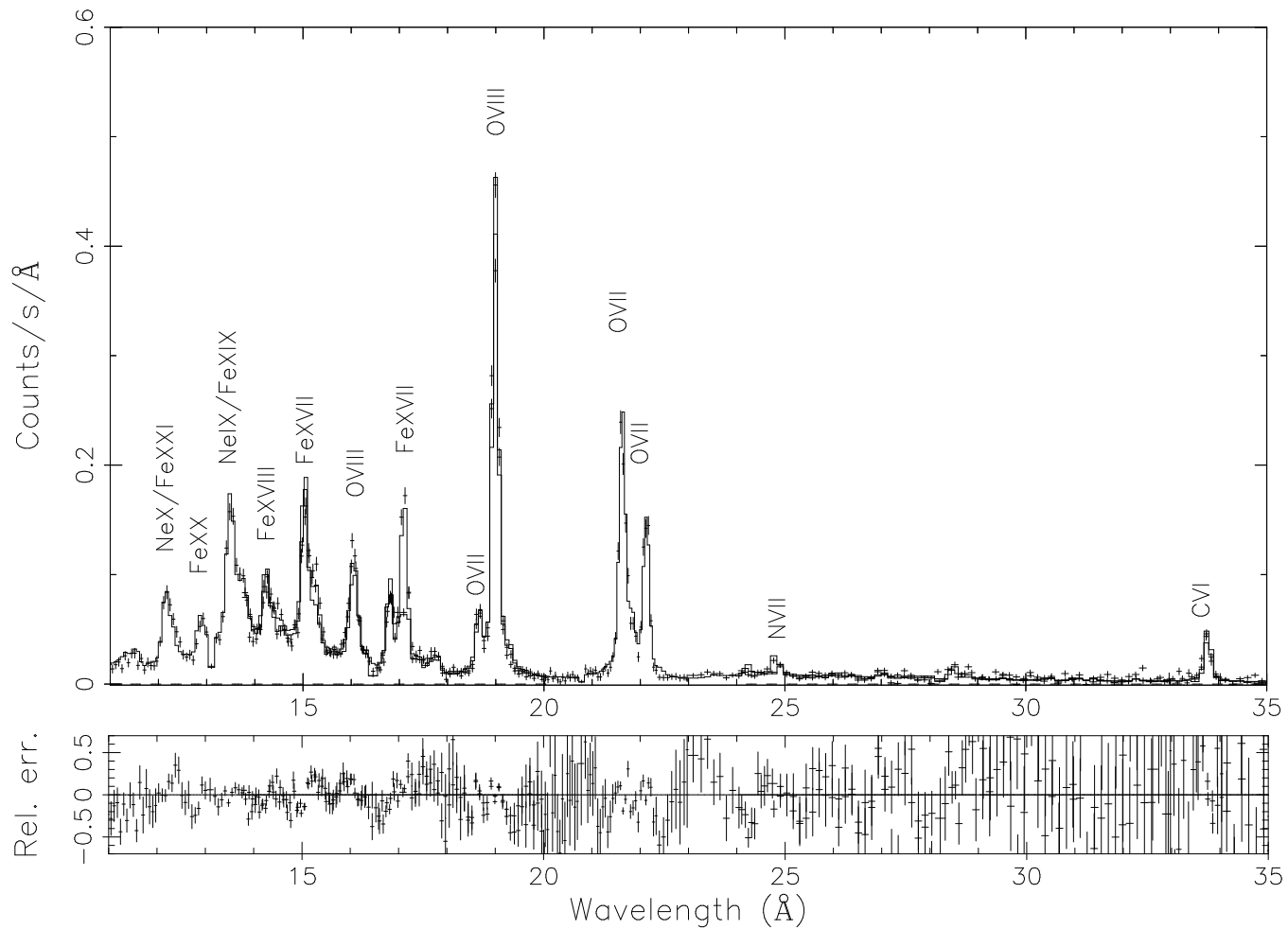}}
\caption[]{The first order \rgs\ spectra of \dem\ in the range 11--35 \AA. \rgs 
-1 and \rgs -2 spectra are overplotted. The
solid line represents a best-fit 2 component NEI model ($kT_{\rm e}=$0.2
\& 0.8\ keV). The most prominent line blends are labeled.}
\label{fig:rgs_fit} \end{figure*}

The first order \rgs\ spectrum of \dem\ is shown in Fig.~\ref{fig:rgs_fit}. The
spectrum is dominated by emission lines, the most prominent of which are
from transitions of highly ionised ions of C, O, Fe and Ne. The
\ion{O}{viii}-L$\alpha$ and \ion{O}{vii}-triplet lines are particularly
strong. Weaker, but unambiguous, emission from ions of N, Mg and Si are
also present. While the second order \rgs\ spectrum has higher spectral 
resolution, it does not provide any additional information. The second 
order spectrum is also of lower statistical quality and was therefore not used. 

We fitted the \rgs\ 1\&2 spectra, extracted from the entire source, 
simultaneously. A response matrix
appropriate to the spatial extent of the source was generated as follows. A
spatial mask corresponding to the \rgs\ aperture was imposed on the \chandra\
ACIS image, and the intensity distribution was integrated over the \rgs\
cross-dispersion direction. The resulting profile was convolved with the
\rgs\ point source response matrix, generated with the SAS task RGSRMFGEN. We
find that one plasma component does not describe the data sufficiently well
and that at least two NEI plasma components are needed for obtaining a good
fit to the \rgs\ spectrum. A hotter component ($kT_{\rm e} \sim 0.8$ keV) 
describes the
spectrum below $\sim 20$ \AA, while a cooler component ($kT_{\rm e} \sim 0.2$ 
keV) is needed to
model the emission from the \ion{O}{vii}-triplet through \ion{C}{vi}. When 
fitting the abundances we used O as the
reference atom and, for technical reasons, kept its abundance fixed to solar.
We use O because it has the strongest emission and would thus give the least
uncertainty in the abundance determination. We allowed the abundance of H, C,
N, Ne, Mg, Si and Fe to vary with respect to O. The He abundance was pegged
to the H value. The best fit results, given in Table~\ref{tab:tab1}, gives a
${\chi}^{2}/d.o.f \sim 1869/945$. We obtain a best fit $N_{\rm
H} = (9.5{\pm 3.1}){\times}10^{24}$\ m$^{-2}$ and a systematic redshift of $380 
{\pm 120}$\ 
kms$^{-1}$, which is consistent with the LMC radial velocity (278 kms$^{-1}$). 
 The derived elemental abundances are consistent with the average LMC abundance
values. We discuss this more extensively in Sec. 4.2.

The model provides very good fits to the data, though a few discrepancies 
exist. For the most
part, the fit discrepancies in Fig.~\ref{fig:rgs_fit} are a result of the 
difference between the broad band spatial profile used in the
response and the actual spatial profiles of individual emission lines.
The images in Fig.~\ref{fig:rgs_O} and the monochromatic images in
Sec.~\ref{sec:monochromatic} show that the remnant's morphology depends
strongly on wavelength. 

\begin{figure}
\resizebox{\hsize}{!}{\includegraphics[angle=-90]{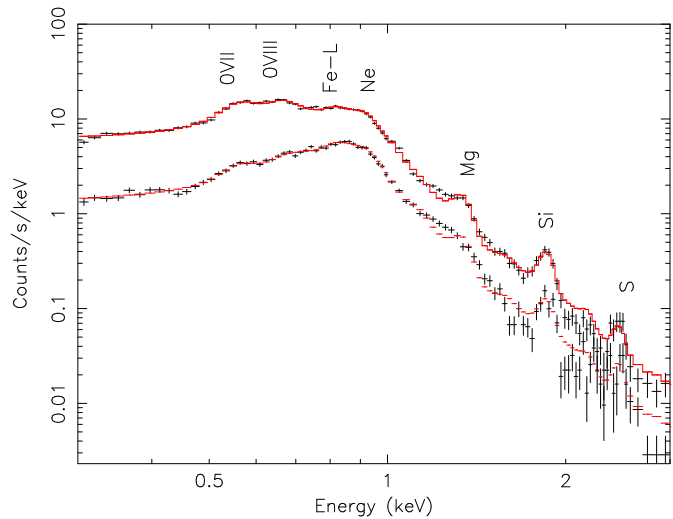}}
\caption[]{The \dem\ spectrum obtained with the \pn. The upper curve
represents the spectrum from the shell, while the lower curve is from
the central region. The solid line represents a best-fit NEI models. The most 
prominent line blends are labeled.}
\label{fig:rim_cen} \end{figure} 

In addition to the spectral features which are present in the \rgs\ 
spectra, the \epic\ spectra also show more prominent emission from Si and also 
S. We do not detect any significant Fe-K emission nor emission from Ca and Ar. 
We extracted and fitted the \pn\ spectrum from the entire source. We use the 
\pn\
spectra because they provide the best statistics. As a starting point we 
used the same model and parameters obtained from the \rgs\ fits. We first 
allow only the normalisations of the plasma components to vary and 
subsequently we allowed the absorption, temperature, ionisation age and 
abundances to vary. The fit results are supplied in Table~\ref{tab:tab1}. The 
derived parameters, except the normalisations, from the fit to the \pn\ data 
remain similar to those obtained from the \rgs. The larger normalisations for 
the \pn\ are due to the higher $N_{\rm H}$ and differences in the \rgs\ and \pn\ 
calibrations (of the order ~10\%).

We also extracted and fitted the \pn\ spectra from both the outer and central
region (see Figs.~\ref{fig:pn_im}~{\&}~\ref{fig:rim_cen}). The spectrum from
the inner region is dominated by a broad emission peak around 0.9\ keV from
Fe-L. The spectrum from the outer rim is softer and shows more pronounced O-K
(\ion{O}{vii} and \ion{O}{viii}) and Mg-K$\alpha$ lines. We fitted the spectrum 
extracted from the rim first. We find that this spectrum can not be well 
described using only one plasma component. A model composed of
two NEI plasma components provides a better fit to the data. The fitting 
procedure was the same as for the full \pn\ spectrum. We then fitted the
spectrum from the inner region. As a model we used one NEI component to account
for emission from the central shell-like structure. In addition to this, we
added the same plasma components and parameters as derived from the fits to the
outer region. Only the normalisations from these components were allowed to
vary. We do this to account for any projected fore/background shell emission.

The results of the fits are supplied in Table~\ref{tab:pnfit}. The shell 
components ($kT_{\rm e} \sim 0.3 \& 0.8$ keV) have
lower temperatures than the central component ( $kT_{\rm e} \sim 1.1$).
Another significant difference between the two regions is the Fe/O \& Si/O 
abundance
ratio. The outer region has a Fe/O$\sim 1.6$ and Si/O$\sim 1.5$, while the 
inner regions has a
much higher Fe/O$\sim 5.9$ and Si/O$\sim 2.3$. The interpretation is discussed 
in Sec. 4.3.

\begin{table*} 
\caption{Best fit parameters for fits to the \rgs\ and \pn\ spectra, extracted 
from the entire remnant. The abundance values are all normalised and relative to 
solar (Anders \& Grevesse \cite{anders}). Note that C and N emission can not be 
determined from the \pn\ data, while the S-K emission is too weak to be detected 
by the \rgs.} 
\label{tab:tab1}
\centerline{ 
\begin{tabular}{|l|c|c|c|c|} 
\hline 
Parametres & \multicolumn{2}{c|}{\rgs} & \multicolumn{2}{c|}{\pn} \\ \hline
 & Comp. 1 & Comp. 2 & Comp. 1 & Comp. 2 \\ \hline \hline
$n_{\rm e}n_{\rm H}V$ (10$^{64}$ m$^{-3}$) & 16.78$\pm$0.71 & 3.51$\pm$0.70 & 
30.10$\pm$10.01 & 5.34$\pm$3.03 \\
$kT_{\rm e}$ (keV) & 0.19$\pm$0.02 & 
0.83$^{+0.05}_{-0.20}$ & 0.21 $\pm$0.04 & 0.82$^{+0.40}_{-0.21}$ \\
$n_{\rm e}t$ (10$^{16}$\ m$^{-3}$s) & 81.10$\pm$20.10 & 7.31$\pm$3.11 & 
81.01$^{+80.10}_{-30.03}$& 7.20$^{+10.10}_{-4.20}$ \\ \cline{2-5}
$N_{\rm H}$ ($10^{24}$\ m$^{-2}$) & \multicolumn{2}{c|}{9.50$\pm$3.10}& 
\multicolumn{2}{c|}{12.10$\pm$6.30} \\ \hline 
\multicolumn{5}{|l|}{Abundances (wrt solar):} \\ \hline
H/O & \multicolumn{2}{c|}{4.96$\pm$ 1.11 } & \multicolumn{2}{c|}{5.0$\pm$2.12} 
\\ 
C/O & \multicolumn{2}{c|}{ 2.3$\pm$ 0.82} & \multicolumn{2}{c|}{-} \\ 
N/O & \multicolumn{2}{c|}{ 0.35$\pm$ 0.21} & \multicolumn{2}{c|}{-} \\ 
Ne/O & \multicolumn{2}{c|}{ 1.50$\pm$ 0.41} & \multicolumn{2}{c|}{1.49$\pm$1.02} 
\\ 
Mg/O & \multicolumn{2}{c|}{ 2.03$\pm$ 1.01} & \multicolumn{2}{c|}{2.11$\pm$1.32} 
\\ 
Si/O & \multicolumn{2}{c|}{ 1.78$\pm$ 1.12} & \multicolumn{2}{c|}{1.96$\pm$1.41} 
\\ 
S/O & \multicolumn{2}{c|}{-} & \multicolumn{2}{c|}{2.01$\pm$1.6} \\ 
Fe/O & \multicolumn{2}{c|}{ 2.65$\pm$ 0.42} & \multicolumn{2}{c|}{2.64$\pm$1.41} 
\\ \hline
\end{tabular}
}
\end{table*} 
\begin{table}
\caption{\pn\ spectral fitting results for the shell and central region. The 
abundance values are all normalised and realtive to solar (Anders \& Grevesse 
\cite{anders}).} 
\label{tab:pnfit} 
\centerline{ 
\begin{tabular}{|l|c|c|c|} 
\hline 
Parametres & \multicolumn{2}{c|}{Shell} & Central region \\ \cline{2-3}
 & Comp. 1 & Comp. 2 & \\ \hline \hline
$n_{\rm e}n_{\rm H}V$ (10$^{64}$m$^{-3}$) & 6.11$_{-0.31}^{+2.04}$ & 
4.52$_{-0.22}^{+1.10}$& 0.65 $\pm 0.22$
\\
$kT_{\rm e}$ (keV) & 0.36$_{-1.34}^{+0.12}$ & 0.79${\pm 0.05}$ & 
1.12$_{-0.33}^{+0.11}$ \\
$n_{\rm e}t$ (10$^{16}$\ m$^{-3}$s) & 4.7$_{-0.5}^{+30.1}$ & 
11.1${\pm 0.22}$ & 4.1${\pm 0.8}$ \\ \hline
\multicolumn{4}{|l|}{Abundances (wrt solar):} \\ \hline
Si/O 		& \multicolumn{2}{c|}{1.53$\pm 0.22$} & 2.32$\pm 0.91$ \\
Fe/O		& \multicolumn{2}{c|}{1.61$\pm 0.34$} & 5.94$\pm 2.10$ \\ \hline
\end{tabular}
} 
\end{table}

Some systematic fit residuals are present in the \pn\ fits. The most prominent 
residual is the underestimation of the spectra at $\approx 1.2$\ keV. This is 
a known problem and is probably due to missing high excitation lines of
\ion{Fe}{xvii-xix} in the plasma code (see Brickhouse et al.\cite{brickhouse}).

\subsection{Monochromatic images and single ion spectroscopy 
\label{sec:monochromatic}}

\begin{figure}
\resizebox{\hsize}{!}{\includegraphics{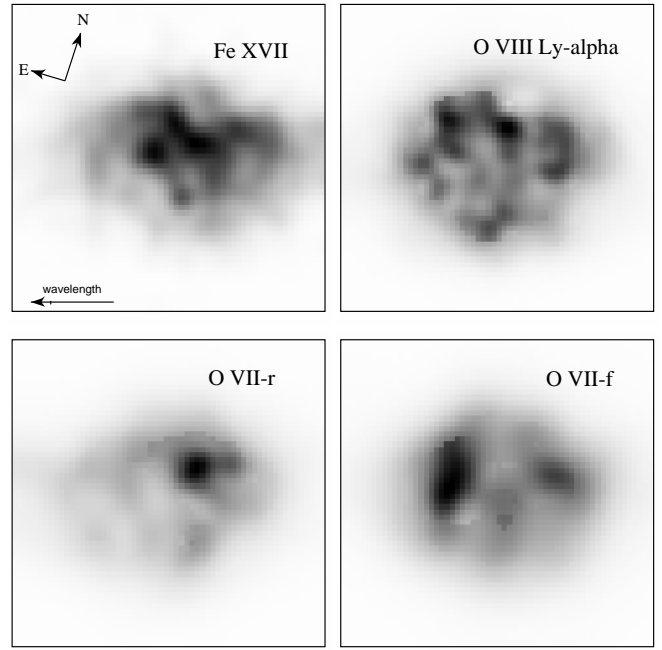}}
\caption[]{\rgs\ monochromatic images for \ion{Fe}{xvii} (at 17.05 \AA) ( 
(upper left)
\ion{O}{viii}~Ly$\alpha$ (upper right), \ion{O}{vii}-resonance (lower left) and 
\ion{O}{vii}-forbidden 
(lower right) lines. The images are plotted with the dispersion and 
cross-dispersion on the X \& Y axis respectively. North is 18\degr 
clockwise, as indicated. The \rgs\ dispersion direction is such that wavelength 
increases from right to left in the images.}
\label{fig:rgs_O}
\end{figure}

We extracted monochromatic images for various lines to probe variations in
temperature and ionisation ages over the remnant and to further investigate the
morphological differences between O and Fe emission. In order to compare \rgs\
line images to one another one has to take into account an important systematic
effect, which is the chromatic magnification that makes long wavelength images
more 'squashed' than the shorter wavelength images. We correct for this effect,
to first order, by using the grating equations. The extent of the source 
of the source produces a maximum broadening of $\pm 0.08$ \AA\ (\ion{O}{vii}-r 
line), equivalent to a Doppler shift in either direction of 1100 kms$^{-1}$. For 
comparison, the intrinsic wavelength resolition of the \rgs\ is $\sim 0.06$ \AA\ 
(FWHM) at 21.6 \AA. However, any Doppler broadening is also convolved along the 
dispersion direction and, depending on its
magnitude, could distort the RGS images. We present images for the
\ion{O}{viii} Ly-$\alpha$, \ion{O}{vii}-resonance (\textit{r}) and
\ion{O}{vii}-forbidden (\textit{f}) line in Fig.~\ref{fig:rgs_O}. We choose
these lines because they are the strongest features in the spectrum and because
the \ion{O}{vii} line ratios are valuable plasma diagnostics. We also extracted
an \ion{Fe}{xvii} image for the purpose of investigating the Fe emission
morphology. The images are all on a on $2{\times}2$\arcmin size. The images
have been smoothed with a gaussian filter (FWHM of 4\arcsec for 
\ion{O}{viii} Ly-$\alpha$ and 8\arcsec for the rest).

\begin{figure}
\resizebox{\hsize}{!}{\includegraphics[angle=-90]{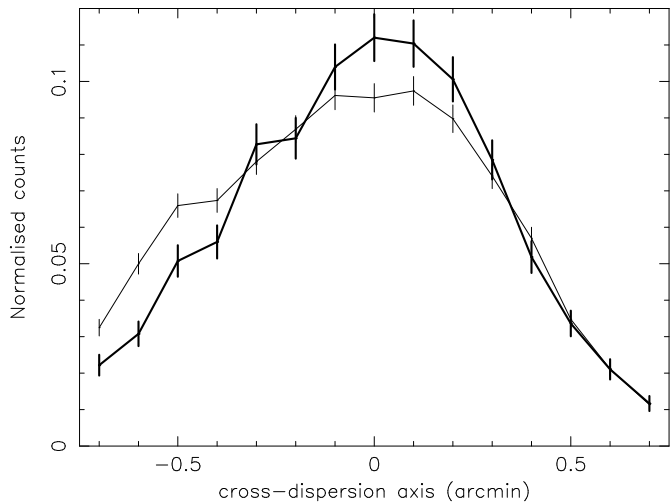}}
\caption[]{\rgs\ cross-dispersion profiles for 
\ion{O}{viii}~Ly$\alpha$ (thin solid curve) and \ion{Fe}{xvii} (thick solid 
curve).}
\label{fig:xdsp_ofe}
\end{figure}

First we turn our attention to the Fe/O emission morphology. The \chandra\
images (Hughes et al. \cite{hughes03}) suggest that most of the O emission 
originates 
from the outer
shell, while the central region is (as compared to the rim) Fe rich.
The \chandra\ observations, however, cannot resolve line blends so these
images are possibly contaminated with underlying continuum emission as
well as emission from different ion species (e.g. Ne). For this reason we turn
our attention to the monochromatic images supplied by the \rgs. The
\rgs\ O images and the \ion{Fe}{xvii} image (see Fig.~\ref{fig:rgs_O}) confirm 
the \chandra\ results. The \ion{Fe}{xvii} image shows a central filled 
morphology with some emission from the outer region.
The O images show more emission from the outer rim with less emission in
the central region. We also examined the cross-dispersion profiles,
displayed in Fig.~\ref{fig:xdsp_ofe}, to ensure the statistical
significance of these results. We use the cross-dispersion since we
have no velocity distortion in this plane, as opposed to the dispersion
direction. Here, again, we see that the Fe emission profile is more
centrally peaked as compared to the O emission.

\begin{figure}
\resizebox{\hsize}{!}{\includegraphics[angle=-90]{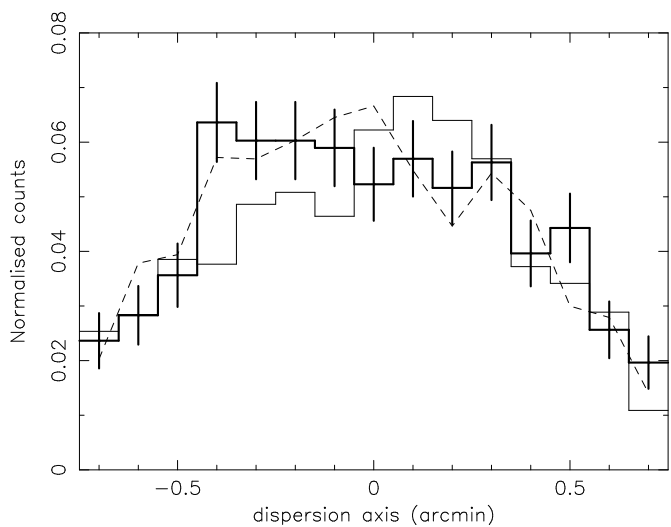}}
\caption[]{\rgs\ dispersion profiles for emission lines of 
\ion{O}{viii}~Ly$\alpha$ 
(dotted curve),
\ion{O}{vii}-resonance (thin solid curve) and \ion{O}{vii}-forbidden 
(thick solid curve). Error bars are provided for \ion{O}{vii}-forbidden line. 
Errors
are typically 50\% and 70\% smaller for the for the
\ion{O}{viii}~Ly$\alpha$ and \ion{O}{vii}-resonance lines, respectively.} 
\label{fig:dsp_O}
\end{figure}

Next we turn our attention to the differences between the morphology of
the \ion{O}{viii}~Ly-$\alpha$, \ion{O}{vii}--$r$ and \ion{O}{vii}--$f$
emission. The images show significant differences between the three
lines. The \ion{O}{viii}~Ly-$\alpha$ emission predominantly comes from
the rim. The \ion{O}{vii}--$r$ emission shows a brighter emission spot in
the northwestern region while the \ion{O}{vii}--$f$ line shows a bright
emission spot on the East rim. To ensure the statistical significance
of these results we consider the dispersion profiles, displayed
in Fig.~\ref{fig:dsp_O}. We now use the dispersion profiles because the
differences in the lines show up more clearly in this plane. These
lines should originate from the same location and have the same
velocity structure since they are from the same ion. Any velocity distortion 
between these images
should thus be insignificant as compared to the resolution of our instrument.
These profiles confirm the variations in the \ion{O}{vii}--$f$ and 
\ion{O}{vii}--$r$ images.

One concern could be the possible contamination of \ion{O}{vii}--$r$ by its 
neigbouring \ion{O}{vii}--$i$ line. Due to the extent of the source, the 
\ion{O}{vii}--$i$ line emission from the West rim could show 
up at the long wavelength shoulder of the \ion{O}{vii}--$r$ emission line (i.e. 
the East rim in the \ion{O}{vii}--$r$ image). 
So any possible contamination would cause an increase in the \ion{O}{vii}--$r$ 
line in the East, we instead see the opposite. The intercombination line 
is, in any case, weak. We checked the spectra from various cross-dispersion 
regions, but did not see any evidence for an enhanced $i$ line.

\begin{figure}
\resizebox{\hsize}{!}{\includegraphics[angle=-90]{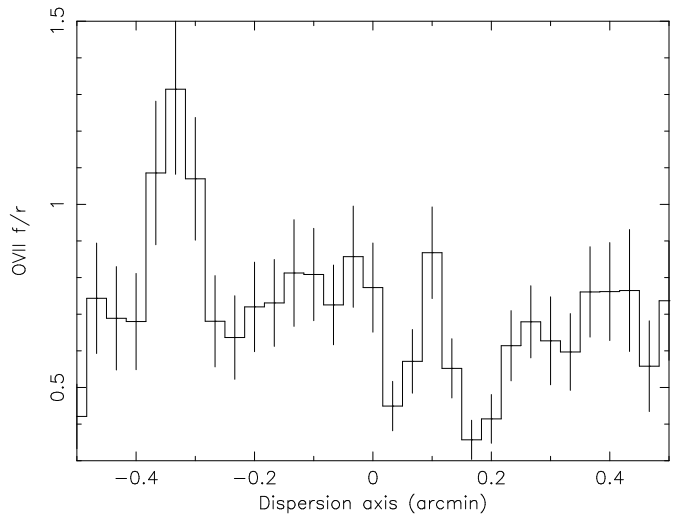}}
\caption[]{\rgs\ \ion{O}{vii} forbidden/resonance line ratio profile in
the dispersion direction.} \label{fig:ratio} 
\end{figure} 

\section{Discussion}

\subsection{Single Ion Spectroscopy}

The \ion{O}{vii} \textit{f}/\textit{r} line ratios, displayed in Fig. 
\ref{fig:ratio}, 
indicate considerable
temperature and ionisation age variations in the outer shell. To further
investigate the implications of the variations in the monochromatic
images we studied the \ion{O}{vii} \textit{f}/\textit{r} ratio. The ratio plot 
has a
mean value of 0.65 and ranges from $\sim$1.3 in the East to as low as
$\sim$0.4 in the northwest. In order to understand the implication of these
ratios we computed the \ion{O}{vii} \textit{f}/\textit{r} for a NEI
plasma using numerous combinations of temperatures and ionisation ages. The 
average ratio of $\sim$0.65 corresponds to a plasma in the temperature range
$kT_{\rm e} \sim 0.1-0.4$\ keV with $n_{\rm e}t > 3{\times}10^{17}$\
m$^{-3}$s. The ratio of $\sim$0.4 in the northwest indicates an ionisation age 
and
temperature combination ranging from ($n_{\rm e}t{\sim}1{\times}10^{16}$\ 
m$^{-3}$s,$kT_{\rm e} \sim 0.3$\ keV)
to ($n_{\rm e}t {\sim}1{\times}10^{17}$\ m$^{-3}$s,$kT_{\rm e} \sim 0.6$\ keV).

The high \textit{f}/\textit{r} value in the East could indicate a low 
ionisation
age ($n_{\rm e}t{\sim}1{\times}10^{15}$\ m$^{-3}$s) plasma. However, such a low
ionisation age plasma should produce strong \ion{O}{vii}-intercombination
(\textit{i}) line emission and we do not see any evidence for this in our data.
The observed line ratio can also be obtained if the gas in this region
is slowly cooling down to a temperature of $\sim0.1$\ keV. Such a situation 
would produce a
recombination spectrum with the observed \textit{f}/\textit{r} ratio and
without a particularly strong \ion{O}{vii}-\textit{i} line. The next
question, if we do indeed have a recombining plasma, is the feasibility
of cooling in \dem. Recombination effects would
become notable if there were some energy draining mechanism
operating at the shock front, e.g. particle acceleration (Dorfi \& 
B{\"o}hringher~\cite{dorfi})
or radiative shocks. If we assume an initial explosion energy of 
$E_{0}=1{\times}10^{44}$ J and an electron density $n_{\rm e} \sim 
3.2\times10^{6}$\ m$^{-3}$, as obtained in Sect.~\ref{mass}, then we estimate 
(according to Falle~\cite{falle}) 
that the radiative cooling phase could commence at an age of $\sim 
1.4{\times}10^{4}$ yrs and this time could be reduced if the blastwave 
encounters higher density ISM material in this region. If we consider that the 
estimated age of \dem\ is $1{\times}10^{4}$ yrs 
then this, together with the fact that radiative shocks have also been observed 
by Ghavamian et al.~\cite{ghavamian}, makes it possible that we have detected 
the effects of recombination. We 
attempted to model the \pn\ spectrum from
the East rim with a recombination model. This model gave a good
description of the data, but required a Fe abundance of greater than 20
times solar while the other elements had sub-solar abundances. We
regard this model as being unphysical. It is probably incorrect to
model emission from the entire rim with a recombination model since
recombination could only be occurring at a localised region(s) along the rim,
as suggested by the bright \textit{f} line emission spot in
Fig.~\ref{fig:rgs_O} and the results of the Fabry-Perot observations of
Ghavamian et al.~(\cite{ghavamian}).

Alternatively, an increase in the $f/r$ line ratio could be introduced by the
presence of resonance scattering of the \ion{O}{vii}--$r$ photons. In
the case of resonance scattering the effective velocity broadening needed to
calculate the optical depth is determined from the maximum differential velocity
over the region with optical depth $\tau = 1$. We estimate a differential 
velocity of 30 kms$^{-1}$, given an overall expansion velocity of 800 
kms$^{-1}$. This differential velocity includes both the gradient in the radial 
velocity component of the shocked gas as well as the effective azimuthal 
velocity difference due to the finite (azimuthal) extent of the last scattering 
surface $\tau = 1$. If we assume a uniform spherical shell of thickness $R_{\rm 
s}/12$ and a velocity broadening of 30 kms$^{-1}$, then we could expect an 
optical depth of $\tau \sim 2.2$ and $\tau \sim 1.0$ in the \ion{O}{vii}-r and
\ion{O}{viii}~Ly$\alpha$ lines, respectively. Following the arguments put 
forward by
Kaastra \& Mewe (\cite{km95}), an optical depth of $\tau \sim 2.2$ could produce
a factor of 2 reduction in the \ion{O}{vii}--$r$ line, which would then produce
the observed $f/r$ ratio. If the \ion{O}{vii} emission originated in a
homogeneously filled spherical shell then one would expect that, in the presence
of resonance scattering, the \ion{O}{vii}--$r$ line image would be more
centrally peaked as the photons would escape along the path of shortest optical
length. This morphology is not observed. However, the effects of resonance
scattering are highly dependent on the morphology and geometry of the system and
resonance scattering could be limited to localised regions. The observed $f/r$
ratio in the East could come about if the shock front encountered a dense
inter-stellar medium (ISM) clump there. The \ion{O}{vii}--$r$ photons would
then be scattered out of the line of sight, thus producing the observed $f/r$
ratio. We neglected the broadening of the line profile due to microturbulence
in the estimation of the optical depth. Microturbulence velocities of a few
hundred kms$^{-1}$ could reduce the optical depth quite considerably. However,
such high microturbulent velocities in \dem\ are unlikely.

Further evidence for resonance scattering is provided by the \ion{O} line 
profiles in Fig.~\ref{fig:dsp_O}. Since the estimated optical depth in 
the \ion{O}{viii}~Ly$\alpha$ line is unity, the expected reduction in this line 
is a factor $\sim$3 less than for \ion{O}{vii}--$r$. The 
\ion{O}{viii}~Ly$\alpha$ and \ion{O}{vii}--$f$ line profiles in 
Fig.~\ref{fig:dsp_O} are quite similar, while most of the deviation is seen in 
\ion{O}{vii}--$r$. This suggests that the high $f/r$ ratio is brought about by a 
reduction in the \ion{O}{vii}--$r$ line rather than an enhancement of the 
\ion{O}{vii}--$f$ lines. This is exactly what we would expect in the presence of 
resonance scattering.

\subsection{Abundances \& Morphology}

\begin{figure}
\resizebox{\hsize}{!}{\includegraphics[angle=-90]{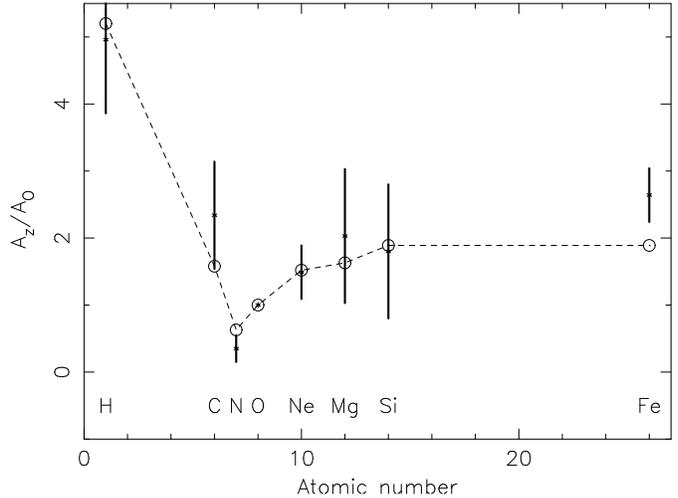}}
\caption[]{Elemental abundance ratios as derived from fits to the \rgs\ 
data. The derived abundance values (crosses) are compared to the 
average LMC abundance values (open circles). The ratios are all relative to 
solar (Anders \& Grevesse \cite{anders}) and normalised to O. The 
LMC values are from Hughes et al. (\cite{hughes2}).} 
\label{fig:abu}
\end{figure}

Fig. \ref{fig:abu} displays the elemental abundance ratios as derived from fits
to the \rgs\ data. The abundances are all relative to solar (Anders \&
Grevesse~\cite{anders}) and normalised to O. The derived abundances are
also compared to the average LMC ratios (Hughes et al. \cite{hughes2}).
The plot shows that the observed abundance values are in good agreement
with the LMC values. Most of the emission seen by the \rgs\ arises from
the shell. This is clear from the morphology of the monochromatic line
images. The shell also dominates the emission in the \epic\ images.
The outer shell is thus interpreted as being emission from a blast wave
which is sweeping up and shock heating ISM material. The idea that the
rim emission is from a blast wave heated ISM is supported by the fact
that the optical spectrum of \dem\ is Balmer-dominated.
Balmer-dominated optical spectra can be interpreted well in the context
of a model in which a high-velocity interstellar shock overtakes at
least partially neutral interstellar gas (Chevalier \& Raymond 1978).
Analysis of the high spatially resolved \chandra\ ACIS-S CCD image shows
that the X-ray outer rim matches the optical H$\alpha$ image nearly
perfectly (Hughes~\cite{hughes01}).

The inner region is Fe rich. This is evident from the fits to the \pn\
spectra which give Fe/O$\sim 1.6$ and Fe/O$\sim 6.0$ for the shell and
inner regions respectively. This Fe enrichment is also visible in the
\rgs\ results, which show a higher Fe/O ratio as compared to the LMC
value. The higher Fe abundance has also been noted in previous ASCA
observations (Hughes et al. \cite{hughes2}). This indicates that the
inner shell structure seen in the \chandra\ images is (reverse shock
heated) Fe-rich SN ejecta. Fe-rich SN ejecta in the inner regions of
SNRs is typical for remnants produced by type Ia SN explosion.

\subsection{Mass estimates \label{mass}}

The results of the line ratios show that the two NEI component model used to fit
the entire spectrum is highly simplified. These temperature and
ionisation age variations in the plasma are probably also the reason why
we need at least two components to model the spectra. Despite our
simplified model, we can still use the results of the full spectral fits
to estimate parameters such as the mean electron densities and emitting
masses. We can infer these parameters from the emission measures, but
to do this we need to make a volume estimate. The \chandra\ images show
that the X-ray emission from both regions originate from limb brightened
shells. The inner shell has an average radius of $\sim$23\arcsec\ while
the outer shell has an average radius of $\sim$37\arcsec. We assume
that the outer shell has a thickness of R$_{\rm s}$/12.
We use the radius and shell thickness assumption to estimate the volume,
which in turn is used to calculate the electron density and masses. We
estimate the outer shell to have an electron density $n_{\rm e} =
3.2\times10^{6}/\sqrt{f}$\ m$^{-3}$ which yields a swept up mass of
M$_{shell}\sim 80{\rm M}_{\sun}{\sqrt{f}}$, where $f$ is the (shell) volume 
filling factor.

The mass estimate of the ejecta (central) region depends on the amount of H 
mixed
into the ejecta during its evolution (see Hughes et al. \cite{hughes03}
for complete explanation). We thus, following the example of Hughes et al. 
(\cite{hughes03}),
compute the masses for two astrophysically plausible scenarios; 
1) We assume that metals are the sole source of electrons 2) assume that a
comparable amount (in mass) of hydrogen has been mixed into the metal rich 
ejecta.
We use the volume estimate for a spherically symetric shell. The \chandra\
images indicate an average radius of $\sim$23\arcsec\
and a shell thickness of $\sim$11\arcsec\ for the central region. 
A pure metal composition yields 1.1 $M_{\sun}$ of Fe and 0.15 $M_{\sun}$ of Si,
while an admixture of H yields 0.7 $M_{\sun}$ of Fe and 0.1 $M_{\sun}$ of Si.
These values are similar to and confirm the results of Hughes et al. 
(\cite{hughes03}). 

There are observational and theoretical indications that type Ia SNe are 
thermonuclear
explosion of accreting white dwarfs (e.g. Nomoto et al. \cite{nomoto2}), 
however, the
exact binary evolution has not been identified. Nomoto et al. ~(\cite{nomoto})
provided some constraints on the progenitor system from the viewpoint of
nucleosynthesis. They presented nucleosynthesis results for various 
deflagration
speeds to constrain the rate of accretion from the companion star. The 
measurement of the Si/Fe, S/Fe, Ar/Fe and Ca/Fe mass ratios would be useful to 
distinguish the type Ia models (Nomoto et al.~\cite{nomoto}). We are only able 
to derive mass estimates for Fe and Si. Nomoto et al.~(\cite{nomoto}) predicts 
a Si/Fe mass ratio of $\sim 0.25$ for the classical (W7) type Ia model and 
$\sim 0.44$ for a delayed detonation (WDD2) model. The present data do not 
allow us to discriminate between models. Deeper observations of \dem\ could 
detect S, Ar and Ca emission which would be useful in identifying the type Ia 
SNe model.

\section{Summary \& conclusions}

The high resolution \xmm\ emission line spectra of \dem\ can be fitted
with two NEI plasma components. The \rgs\ spectra are dominated by
emission from the SNR shell, so the two NEI components needed to fit
these spectra indicate a range of temperatures and ionisation ages
present in the shell. The temperatures ($kT_{\rm e} \sim$\ 0.2 \& 0.8
keV) are consistent with the shock temperature ($kT_s =0.11-0.75$\ keV)
obtained from optical results (Smith et al. \cite{smith}). Temperature
and ionisation age variations are also evident from the difference in
morphology between the \ion{O}{vii} \textit{f} \& \textit{r} lines. 
These lines show an anomalously high ratio ($f/r \sim 1.2$) in the 
East. This ratio could imply that in this region the plasma is
cooling and recombining. However, an
alternative, and equally interesting, possibility is that
the variation in $f/r$ ratios is due to resonance scattering,
which would reduce \ion{O}{vii}--$r$ line emission along lines of sight
with a high O VII column density.

The \chandra\ images reveal a double shock structure in \dem. The \pn\
spectral analysis of these two regions indicates that the inner region
($kT_{\rm e} \sim 1.1$ keV) is hotter than the shell ($kT_{\rm e} \sim 0.2-0.8$ 
keV). The abundance ratios of the outer shell is
consistent with the average LMC values, while the inner region shows a much
higher Fe/O abundance ratio. The outer shell has a much higher mass
($\sim 80 M_{\sun}$) compared to the inner shell ($\sim
2M_{\sun}$). The image and spectral analysis of \dem\ thus suggests
that the remnant has a structure consisting of a hot centre surrounded
by a cooler shell. This is in agreement with shock models like Truelove \& 
McKee (\cite{truelove}), which predict a hotter interior. The outer rim 
represents a blast wave moving out
into the ISM, while the interior emission is from reverse-shock-heated Fe-rich 
stellar ejecta material. The morphology, mass estimates and abundances strongly
suggest that \dem\ is the result of a type Ia explosion as previously
thought. Our analysis also confirms the recent results of Hughes et al. 
(\cite{hughes03}).

\dem\ is a good candidate object to study in order to discriminate between
the various type Ia supernovae models (e.g. Nomoto et al \cite{nomoto}). 
However, the current data do not allow for this. Deeper
observations would provide better statistics to study the ejecta material
in more detail, particularly the density profile of the ejecta 
plasma and the possible detection of Ar, Ca and Fe-K emission.

\begin{acknowledgements}
We thank Ehud Behar for the valuable discussions we had. We also thank the 
referee J. Ballet for his detailed comments.
The results presented are based on observations obtained with XMM-Newton, 
an ESA science 
mission with instruments and contributions directly funded by 
ESA Member States and the USA (NASA). 
JV acknowledges support in the form of the NASA Chandra Postdoctoral 
Fellowship grant nr. PF0-10011, awarded by the Chandra X-ray Center. SRON is 
supported financially by NWO, the Netherlands Organisation for Scientific 
Research.
\end{acknowledgements}

\end{document}